\documentclass[preprint,5p,times,twocolumn]{elsarticle}

\usepackage[utf8]{inputenc}
\usepackage{cite}
\usepackage{amsmath, amssymb, mathrsfs}

\usepackage{array}
\usepackage{enumitem}
\usepackage{comment}
\usepackage{dirtytalk}
\biboptions{sort&compress}

\usepackage{adjustbox}
\usepackage{multirow}

\newcolumntype{C}[1]{>{\centering\let\newline\\\arraybackslash}m{#1}}

\newlength\matlabfigurewidth
\usepackage{subcaption}

\usepackage{booktabs, colortbl, array}
\usepackage{pgfplotstable}
\pgfplotsset{compat=1.8}

\definecolor{rulecolor}{RGB}{0,71,171}
\definecolor{tableheadcolor}{gray}{0.92}

\newcommand\mydots{\hbox to 1em{.\hss.\hss.}}
\newcommand\seq[3]{\ensuremath{[#1]_{#2}^{(#3)}}}

\graphicspath{{./fig}}

\begin{document}

\begin{frontmatter}

\title{Scheduling Power-Intensive Operations of Battery Energy Storage Systems and Application to Hybrid Hydropower Plants}

\author[inst1]{Stefano Cassano}
\author[inst1]{Fabrizio Sossan}

\affiliation[inst1]{organization={University of Applied Sciences of Western Switzerland (HES-SO)},
            addressline={Rue de l'Industrie 21}, 
            city={Sion},
            postcode={1950},
            country={Switzerland}}

\begin{abstract}
This paper proposes a novel set of power constraints for Battery Energy Storage Systems (BESSs), referred to as Dynamic Power Constraints (DPCs), that account for the voltage and current limits of the BESS as a function of its State of Charge (SOC). 
These constraints are formulated for integration into optimization-based BESS scheduling problems, providing a significant improvement over traditional static constraints.
It is shown that, under mild assumptions typically verified during practical operations, DPCs can be expressed as a linear function of the BESS power, thus making it possible to retrofit existing scheduling problems without altering their tractability property (i.e., convexity).
The DCPs unify voltage and current constraints into a single framework, filling a gap between simplified models used in BESS schedulers and more advanced models in real-time controllers and Battery Management Systems (BMSs).
By improving the representation of the BESS's power capability, the proposed constraints enable schedulers to make more reliable and feasible decision, especially in power-intensive applications where the BESS operates near its rated power.
To demonstrate the effectiveness of the DPCs, a simulation-based performance evaluation is conducted using a hybrid system comprising a 230 MW Hydropower Plant (HPP) and a 750 kVA/500 kWh BESS. Compared to state-of-the-art formulations such as static power constraints and DPC formulations without voltage constraints the proposed method reduces BESS constraint violations by 93\% during real-time operations.

\end{abstract}


\end{frontmatter}


\section{Introduction}

Battery Energy Storage Systems (BESSs) attract significant interest in electrical power systems thanks to maturing electrochemistry and decreasing costs. BESS applications range from energy arbitrage and self-consumption \citep{BARVA2024110561, 8783608, BENNETT2015122} to grid frequency support \citep{ALSHARIF2023107630, OKAFOR20233681, ZHANG2025124541}, hybridization of power plants \citep{HOFKES2024111247, VAZQUEZRODRIGUEZ2023108782,BHATTI2023120894}, and non-wire grid reinforcements \citep{KHAN2024112373,DELAVARI2024110407, gupta2023optimal} in both front- and behind-the-meter settings.

Operating a BESS requires two hierarchical software layers operating at different temporal scales: a real-time controller and a scheduler. The real-time controller, together with the battery management system (BMS), ensures that the charging/discharging power and reactive power setpoints requested to the BESS respect the physical limits of the system, ensuring safe operations. On the AC side, BESS power limits are modeled by the capability curve of the power converter (e.g., \citep{nick2014optimal,MEHRJERDI2019810}), which varies as a function of the voltage magnitude of the grid connection point and of the battery DC bus \citep{zecchino2021optimal}. The power limits on the AC side are proxies to the voltage and current constraints on the DC side, which ultimately represent the fundamental physical constraints of the system, and those that are carefully monitored by the real-time controller and BMS. Assessing the feasibility of power setpoints with respect to the voltage and current constraints of the DC bus is a central task of the real-time controller and BMS; it can be performed with two main classes of methods: characteristic maps and model-based approaches. The first uses lookup tables to map the maximum current and power of the battery to the State of Charge (SOC), the state of health, the terminal voltage, and the temperature \citep{Sauer:112626,CN101133514B,osti_1186745, KOLTERMANN2023121428}; the second uses equivalent circuit models to establish an analytical relationship among the involved electrical quantities, and includes the SOC-limited \citep{SUN2012378}, voltage-limited ohmic resistance \citep{7122358}, and voltage-limited extrapolation methods \citep{7122358,6314892}.
On a slower time scale than the real-time controller, BESS schedulers ensure that sufficient energy is available to provide the prescribed grid services over time. Energetic needs are modeled with dedicated state-of-energy (or state-of-charge) constraints, supplied with time-series forecasts of the power of the grid services to provide. State-of-the-art schedulers are formulated as a constrained techno-economic optimization problem. Their development has been extensively documented in the literature for various uses, including with strategies to limit battery degradation (e.g., \citep{he2015optimal, namor2016assessment, AMINI2023104492, shi2018convex}).

As the key focus of schedulers is energy, they typically adopt simplified models of the BESS power compared to BMS and real-time controller to ease the resolution of the optimization problem. In the existing literature, a standard approach to model power constraints in scheduling problems is using inequalities in the form of $|B_t| \le B^r$ (\footnote{Or, if the AC reactive power $Q(t)$ is a decision variable of the problem too, the four-quadrant capability curve of the converter $B(t)^2 + Q(t)^2 \le B^{r^2}$ is used as a generalization of the constraint $|B(t)| \le B^r$.}), where $B$ is the discharging power of the BESS in the time interval $t$ and $B^r$ is the rated power of the system, as in \citep{gupta2023optimal,7932132,elsaadany2023battery,en14248480,SELIM202311741,PEESAPATI2024101227,inproceedings1,SONG2024100378} and as documented in the review papers \citep{HANNAN2021103023,HOSSAINLIPU2022132188}. The popularity of these constraints in optimization-based scheduling problems stems from their linearity in the decision variables, which does not alter linearity or convexity of the original optimization problem. However, compared to the models used in real-time control and BMS and introduced above, approximated power constraints will results in innacurate estimations of the dynamic power capability of the BESS. While such inaccuracies may be acceptable in some contexts, this paper shows that in power-intensive applications, namely when the BESS frequently operates near its rated power, these modeling inaccuracies might impact negatively the performance of the scheduler, resulting in unfeasible scheduling actions; it is thus critical to implement more accurate power constraints compared to standard ones. In this context, this paper proposes a set of linear power constraints for BESS scheduling problems that capture the BESS's underlying voltage and current constraints as a function of its SOC with a significantly larger accuracy than existing formulations. We prove that these constraints, under assumptions normally verified in practical operations, are linear in the battery power; this makes it possible to use the proposed constraints to retrofit existing formulations (including for diverse applications and beyond the application example proposed in this paper) without impacting their tractability properties. 

It is important to remark that applying the same class of models discussed above for BMS, such as nonlinear physical models, lookup tables, or even rule-based models, to scheduling problems is generally not pursued as it would impair tractability and affect negatively on problem properties (e.g., resolution time, convergence, lack of global optimum, reproducibility of the solution).

The next paragraphs discuss these contributions compared to the existing research works that have proposed more accurate power constraints for BESS scheduling problems.

The work in \citep{7752949} implements power constraints cognizant of the battery voltage and current limits. However, being the model used for short-term scheduling and real-time control, the dependency on battery SOC is omitted; compared to this work, we model the dependency between the battery open-circuit voltage (OCV) and SOC and show that, proving the linearity of these constraints under mild assumptions.

The work in \citep{8770143} uses an equivalent circuit model to formulate battery power constraints that are aware of the battery current limits. Although current limits are essential, voltage limits are also critical; compared to \citep{8770143}, our formulation includes the voltage limits too: by way of a performance comparison, we show that implementing voltage constraints positively affects performance.

The authors of \citep{kaczorowska} formulate battery power constraints that are dependent on the SOC, deriving curves that are qualitatively similar to those found in this paper. However, these curves are not implemented in a scheduling problem. Compared to \citep{kaczorowska}, we derive a closed-form expression of these curves as a function of the SOC and BESS power and demonstrate their computationally efficient integration into BESS scheduling problems. 

In the context of smart charging scheduling problems for electric vehicles (EVs), the works in \citep{Cao,7244227,9060993} introduces convex constraints for the recharging power that are aware of the current limits of the EV battery. Compared to these works, our contributions extend to considering bidirectional power for grid-connected BESS.

In summary, the key contribution of this work compared to previous research on BESS scheduling problems is a more accurate (linear) formulation of the BESS charging/discharging power constraints based on an improved modeling of the electrical properties of the BESS. As discussed above, most studies on BESS scheduling problems adopt simplified static power constraints, with only a few papers addressing the formulation of more accurate power constraints. In this paper, besides arguing that more accurate power constraints in BESS scheduling problems are essential for power-intensive applications, we propose a novel set of linear power constraints that can be efficiently integrated into BESS scheduling problems that account for both voltage and current constraints, increasing the accuracy of the scheduling action. Because the existing works do not yet document a unified way to handle current and voltage constraints in the scheduling problem as a function of the BESS SOC and with linear power constraints, this stems as the main contribution of this paper compared to the existing literature. This contribution goes toward filling the methodological gap between the models of the BESS power used in BESS schedulers and more advanced ones used in real-time controllers and BMS.

The methodology developed in this paper refers to a lithium-ion BESS. However, it can be adapted to model the power constraints of any energy storage resource that can be modeled with equivalent circuit models; this includes other battery electrochemistries, such as sodium ions, and, possibly, fuel-cell/electrolyzer systems.

The rest of this paper is organized as follows. Section \ref{sec:Problem} presents the formulation of a scheduling problem with conventional $|B_t| \le B^r$ constraints to provide a practical example of when they fall short, and lay the methodological foundation of the paper. Section \ref{sec:DPCs} derives the proposed set of power constraints and discusses their properties. Sections \ref{sec:appl} and \ref{sec:results} present the application example and discuss the results and performance comparison, respectively.

\section{Limitations of static power constraints in scheduling problems}\label{sec:Problem}

\subsection{Formulation of the scheduling problem}
The purpose of a BESS scheduler is to compute an appropriate charging/discharging power trajectory of the BESS to ensure that sufficient power and energy levels are available during real-time operations to the real-time controller to provide the services for which it is designed. E.g., if a battery is nearly fully discharged and discharging power is needed to provide frequency regulation, the scheduler should intercept this need (by way forecasts of the service to provide) and recharge the battery preemptively. BESS schedulers are conveniently implemented as constrained optimization problems, where the cost function models the operational objective to achieve (e.g., peak shaving, electricity cost optimization, provision of frequency control) and the constraints refer to the operational limits of the BESS (e.g., rated power and energy capacity). An example of a BESS scheduler from the literature is the following. Let $\widehat{B}_t$ be the predicted BESS power (positive when discharging, negative when charging) at time $t$ in kW. It can be expressed as:
\begin{align}
    \widehat{B}_t = \widehat{P}_t + F_t \label{eq:spc}
\end{align}
where $\widehat{P}_t$ is a point prediction of the power demand of the service to provide (e.g., primary frequency control, peak shaving), and $F_t$ is a problem slack variable that releases the BESS from providing the prescribed service if this leads to infeasible SOC and power conditions. We refer to the sequence $F_t$ for $t=0,\dots, T$ as \say{offset profile}.

The BESS power should not exceed the power rating of the system. Assuming that the system operates at a unitary power factor, this requirement reads as:
\begin{align}
-B^r \le \widehat{B}_t \le B^r,
\end{align}
where $B^r$ is the kVA rating of the system; we refer to these constraints  as Static Power Constraints (SPC) because they do not depend on time or other quantities. The battery SOC in per unit at time $t+1$ is approximated as a function of the charging/discharging power $\widehat{B}_t$ (in kW) as:
\begin{align}
SOC_{t+1} = SOC_{0} &- \frac{\Delta_{T}}{E^r} \sum_{\tau=0}^{t} \left( \dfrac{1}{\eta}[B_\tau]^+ - \eta [B_\tau]^- \right) \label{eq:soc_0}
\end{align}
where $SOC_{0}$ is a known initial state of charge, $\Delta_T$ is the duration in hour of the time interval $t$, $E^r$ is the BESS rated energy storage capacity in kWh, $\eta$ is the conversion efficiency, and $[\cdot]^+$ and $[\cdot]^-$ denote the positive and negative part of the argument, respectively. For compactness, we denote with the notation $\seq{x}{\underline i}{\overline i}$ a vector composed of all the scalars from $x_{\underline i}$ to $x_{\overline i}$. For example:
\begin{align}
\seq{\widehat{B}}{0}{t} = \begin{bmatrix}
\widehat{B}_0 & \widehat{B}_1 & \dots & \widehat{B}_t
\end{bmatrix}.
\end{align}
With this notation, Eq.~\eqref{eq:soc_0} becomes:
\begin{align}
SOC_{t+1} = SOC_{0} &- h\left(\seq{\widehat{B}}{0}{t}\right) \label{eq:soc},
\end{align}
where $h(\cdot)$ is a scalar-valued function defined as:
\begin{align}
h\left(\seq{x}{\underline i}{\overline i}\right) = \frac{\Delta_{T}}{E^r} \sum_{i=\underline i}^{\overline i} \left( \dfrac{1}{\eta} \left[ x_i \right]^+ - \eta \left[ x_i \right]^- \right). \label{eq:function_h}
\end{align}

With these models in place, the BESS scheduling problem can be stated as finding the sequence $\seq{F}{0}{T}$ while obeying to the constraints on the BESS power and SOC and while requesting to the BESS the power $\widehat{P}_t$ in \eqref{eq:spc}. The scheduler formally reads as the constrained optimization problem in \eqref{eq:scheduler0} that minimizes the norm-2 of the offset profile as a best-effort attempt to provide the required power $\widehat{P}_t$. The cost function includes customizable non-negative weights $c_t$ that can be used to assign more or less importance to the BESS action at a given time (\footnote{Other problem formulations, in particular of the cost function, are possible. This formulation is meant to provide an illustrative example of the limitations of the problem constraints.}). The problem formulation is:
\begin{subequations}\label{eq:scheduler0}
\begin{align}
    \underset{\seq{F}{0}{T}}{\text{arg~min}} \left\{ \sum_{t=1}^{T} c_t F_t^2 \right\}
\end{align}
subject to:
\begin{align}
& \widehat{B}_t = \widehat{P}_t + F_t, && t=0,1 \dots, T\label{eq:c0}\\
& -B^r \le \widehat{B}_t \le B^r, && t=0,1 \dots, T \label{eq:powerconstraint0}\\
& SOC_{t+1} = SOC_0 - h\left(\seq{\widehat{B}}{0}{t}\right), && t=0,1 \dots, T \\
& \underline{SOC} \le SOC_{t} \le \overline{SOC}, && t=1,2 \dots, T+1 \label{eq:soc0}
\end{align}\end{subequations}

\subsection{Limitations of the above-mentioned formulation}\label{sec:limits}
\begin{figure}[ht!]
    \centering
    \includegraphics[width=1\columnwidth]{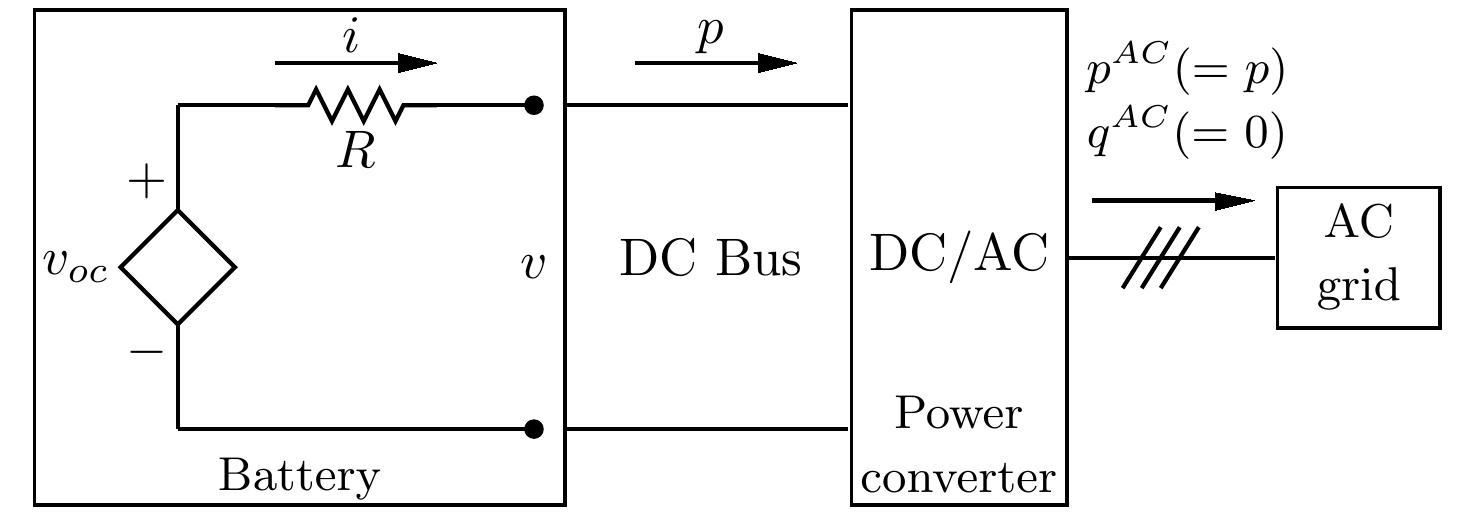}
    \caption{A diagram of a BESS connected to the grid showing the (steady-state) equivalent circuit of a BESS, the DC bus and the DC-to-AC power converter.}
    \label{fig:batt_circuit}
\end{figure}

The power constraints in \eqref{eq:powerconstraint0} are an approximation of the physical constraints of a battery system that stems from assuming a constant battery voltage. To prove this proposition, we resort to the steady-state equivalent circuit model of a BESS, shown in left part of Fig.\ref{fig:batt_circuit}. This circuit models the voltage on the battery DC bus $v$ as a function of the battery discharging current $i$ (positive when discharging, negative when charging). The equivalent circuit consists of the battery open circuit voltage $v_{oc}$ and the series resistance $R$, that combines the internal resistance of the battery and the Thevenin equivalent resistance of the converter. As converter losses are embedded in $R$, the DC-to-AC power conversion stage is assumed ideal (lossless). Converter switching dynamics are neglected because much faster than the time scale of the scheduler. Battery dynamics are also assumed at steady state; battery dynamic models, such as the two-time-constant model, can be however incorporated in the proposed modeling framework. The values of these two circuit elements are BESS-specific and can be estimated from measurements.

With reference to the equivalent circuit model of Fig.~\ref{fig:batt_circuit}, the fundamental physical constraints of the system refer to the voltage $v$ and the current $i$ and are:
\begin{align}
    \underline{v} \le v \le \overline{v} \label{eq:m:vcon}\\
     -\overline{i} \le i \le \overline{i} \label{eq:m:icon}.
\end{align}
where $\underline{v},\overline{v}$ denotes the allowed voltage range, and $\overline{i}$ is the rated current. Assuming the battery voltage as a constant value at the rated value $v^r$ (with $\underline{v}\le v^r \le \overline{v}$) implies that the inequality in \eqref{eq:m:vcon} is verified by construction. In addition, the inequality in the current \eqref{eq:m:icon} can be equivalently rendered as an inequality in the power as follows:
\begin{subequations}
\begin{align}
    -\overline{i} v^r \le i v^r \le \overline{i} v^r \\
    -\overline{p} \le p \le \overline{p}
\end{align}
\end{subequations}
where $p = i v^r$ is the DC power delivered by the circuit when $v=v^r$, and $\overline{p} = \overline{i} v^r$ is the rated power. This reasoning proves that, under the assumption of constant battery voltage, power constraints are a valid proxy to the battery current constraints, justifying their adoption in \eqref{eq:powerconstraint0}. However, because the assumption of constant battery voltage does not hold due to resistive losses and variable open-circuit voltage of the battery (as fully explained in Section~\ref{sec:DPCs}), schedulers that (tacitly) operates under this assumption might cause unfeasible BESS operations, as illustrated in the following example.

\subsection{Motivating example}
It is considered a BESS with a rated power of $B^r=720$~kW and an energy capacity of $E^r=560$~kWh. The initial time interval is $t=0$, the duration of the optimization horizon is $T=5$, and the integration time is $\Delta_t=5$~minutes. The initial battery state of charge is $SOC_0=20$\%. At $t=2$, the BESS is called to provide a discharging power of 600~kW for 5 minutes, amounting to 50~kWh: this information is encoded in the vector $[\widehat{B}]^{(5)}_0$, which is made by all zeros, except for the third element, which is 600~kW. The amount of requested energy corresponds to about 9\% of the battery's nominal energy capacity, assuming ideal discharging efficiency. In addition, the battery state of charge limits are $\underline{SOC}=5\%$ and $\overline{SOC}=95\%$. 

For the sake of illustration and before proceeding to the numerical resolution, it is useful to deduce the solution to the scheduling problem by reasoning: the solution of the unconstrained optimization problem associated with \eqref{eq:scheduler0} is a vector of all zeros, namely $[F]_0^{(5)} = [0]^{(5)}_0$; because this solution does not violate the constraints \eqref{eq:c0}-\eqref{eq:powerconstraint0} (indeed, the discharging power of 600~kW required to the battery is smaller than the converter rated power of 720~kW, and the SOC drop of 9\% applied to the SOC initial value of 20\% is still within the feasible range 5\%-95\%), one can conclude that the all-zero vector is also solution to the constrained optimization problem in \eqref{eq:scheduler0}.

\begin{figure}
	\centering
	\subfloat[\label{fig:ex:a}]{%
		\includegraphics[width=1\columnwidth]{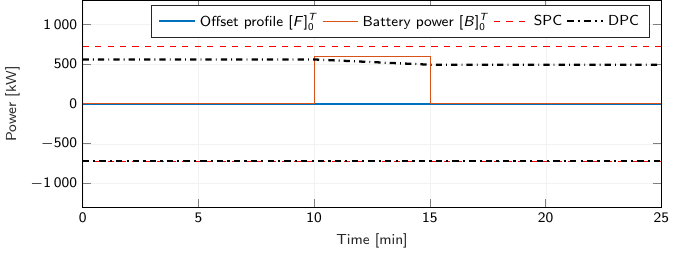}}
	\\
	\subfloat[\label{fig:ex:b}]{%
		\includegraphics[width=1\columnwidth]{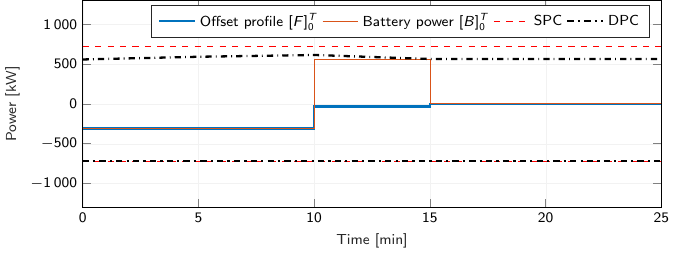}}
	\caption{BESS discharging power and power limits determined by the scheduler under different constraints with Static Power Constraints (SPCs) in (a), and with Dynamic Power Constraints (DPCs) in (b).}
	\label{fig:ex}
\end{figure}

Resolving the scheduling problem in \eqref{eq:scheduler0} numerically yields the same solution deduced by reasoning, as shown in the upper panel of Fig.~\ref{fig:ex}. Fig.~\ref{fig:ex} shows, with red dashed lines, the static converter rating at $\pm 720\text{~kW}$, corresponding to the SPCs, and, with black dashed-dotted lines, the more accurate power limits calculated by considering the elements enounced in \ref{sec:limits} and that will be formalized in the next section. We refer to this second set of constraints as Dynamic Power Constraints (DPCs).

Fig.~\ref{fig:ex} (a) shows that the BESS power respects the SPCs, but it exceeds the upper bound of the DPCs. Two elements cause this violation: first, high power requires high current, making under-voltages more likely owing to large voltage drops on the resistance; second, low SOC values result in low open-circuit voltages, causing large current, reinforcing voltage drops and possibly resulting in over-currents. In summary, delivering high discharging power is more prone to over-voltages and over-currents. Because the standard scheduling problem is not informed about voltage and current limits, its solution does not respect the DPCs. The next section shows the formulation of DPCs for the scheduler that is aware of the voltage and current constraints of the BESS' DC bus.

For completeness, Fig.~\ref{fig:ex} (b) shows the results of applying the scheduler augmented with the proposed constraints: as it can be seen, the BESS discharge limit increases in the first 10 minutes of the scheduling horizon, putting the BESS in a better position to provide the target discharge power of 600~kW. The scheduler achieves this result by preemptively charging the BESS so that it can achieve a higher voltage and respect the constraints.

\section{Dynamic Power Constraints for the BESS Scheduling Problem}\label{sec:DPCs}
This section describes a formulation for accurate power constraints for BESS scheduling problems, called dynamic power constraints (DPCs), that account for the current and voltage physical constraints of the battery. This is the central contribution of the paper.

\subsection{From voltage and current constraints to power constraints}

We are concerned with finding the largest charging (maximum value) and discharging (minimum value) power that a BESS, modeled with the equivalent circuit of Fig.~1, can deliver, subject to voltage and current constraints in \eqref{eq:m:vcon} and \eqref{eq:m:icon}. The battery terminal voltage $v$ in the circuit in Fig.~1 is:
\begin{align}
v = v_{oc}\left(SOC\right) - Ri. \label{eq:v}
\end{align}
where $v_{oc}\left(SOC\right)$ denotes the battery open-circuit voltage as a function of the battery SOC. For compactness and with abuse of notation, we omit the dependency of $v_{oc}$ on $SOC$ in the following formulation.
The power $p$ delivered at the DC circuit terminals is:
\begin{align}
p = vi = (v_{oc} - Ri) i = v_{oc}i - Ri^2. \label{eq:dcpower}
\end{align}
From the maximum power transfer theorem, the current $i_{max}$ corresponding to the maximum power transfer is computed by equating the derivative of \eqref{eq:dcpower} to zero:
\begin{align}
\dfrac{dp}{di} = v_{oc} - 2Ri \label{eq:dpdi} \\ 
i_{max} = \dfrac{v_{oc}}{2R}. 
\end{align}
From this it follows that power $p$ increases monotonically with the current for $i \le i_{max}$. Because for real batteries, the rated current is typically much smaller than $i_{max}$, this condition holds and will be assumed as one operative assumption in the rest of this paper.

As introduced in Section~\ref{sec:Problem}, the electrical physical constraints of a battery system consist of the battery voltage and current constraints:
\begin{subequations}
\begin{align}
    \underline{v} \le v \le \overline{v} \label{eq:v_con} \\
    -\overline{i} \le i \le \overline{i} \label{eq:i_con}.
\end{align}
\label{eq:con}
\end{subequations}
Replacing \eqref{eq:v} into \eqref{eq:v_con} gives:
\begin{align}
    \underline{v} \le v_{oc} - R i \le \overline{v} \label{eq:v_con_a}
\end{align}
Rearranging the terms of \eqref{eq:v_con_a} yields to the following inequalities:
\begin{align}
\frac{v_{oc} - \overline{v}}{R} \le i \le \frac{v_{oc} - \underline{v}}{R}. \label{eq:v_con_e}
\end{align}
The left-hand side of \eqref{eq:v_con_e} is by construction the lower bound of the current such that the voltage constraints of the battery are respected; the right-hand side is the upper bound.

According to the operative assumption stated earlier, the battery power monotonically increases with the current; thus, the current bounds in \eqref{eq:v_con_e} can be used to find power bounds; replacing the current lower and upper bounds of \eqref{eq:v_con_e} into \eqref{eq:dcpower} yields the power bounds such that the voltage constraints are respected. For the power upper bound, this reads as:
\begin{subequations}\label{eq:p_bar_voltage}
\begin{align}
\begin{aligned}
\overline{p}_v &= v_{oc}\left(\frac{v_{oc} - \underline{v}}{R}\right) - R \left(\frac{v_{oc} - \underline{v}}{R}\right)^2= \\
&= \dfrac{\underline{v}}{R} \left( v_{oc} - \underline{v} \right).
\end{aligned}
\end{align}
For the lower bound, it is:
\begin{align}
\underline{p}_v = \dfrac{\overline{v}}{R} \left( v_{oc} - \overline{v} \right).
\end{align}\end{subequations}
Similarly to the power bounds due to the voltage constraints, one can compute the power bounds due to the current constraints by replacing the lower and upper bounds of \eqref{eq:i_con} into \eqref{eq:dcpower}. These power bounds read as:
\begin{subequations}\label{eq:p_bar_current}
\begin{align}
\underline{p}_i &= -v_{oc}\overline{i} - R \overline{i}^2 \\
\overline{p}_i &= v_{oc}\overline{i} - R \overline{i}^2.
\end{align}\end{subequations}
In summary, the initial set of constraints on voltage and current in \eqref{eq:con} can be replaced by:
\begin{subequations}
\begin{align}
& p \ge \underline{p} = \text{max} \left(\underline{p}_v, \underline{p}_i\right) \\
& p \le \overline{p} = \text{min} \left(\vphantom{\underline{p}_v}\overline{p}_v, \overline{p}_i\right).
\end{align}
\label{eq:dpc_max}\end{subequations}
The advantage of \eqref{eq:dpc_max} compared to \eqref{eq:con} is that, in BESS scheduling problems, the decision variable is the battery power and not the battery current and voltage; in other words, the reformulation in \eqref{eq:dpc_max} allows to write explicit constraints on the problem decision variable (i.e., the BESS power) that enforce the underlying voltage and current constraints.

The rest of this section concerns implementing the constraints \eqref{eq:dpc_max} in the scheduling problem. The terms $\underline{p}_v, \underline{p}_i, \overline{p}_v, \overline{p}_i$ on the left-hand side of \eqref{eq:dpc_max} are as defined in \eqref{eq:p_bar_voltage} and \eqref{eq:p_bar_current}. In order to calculate \eqref{eq:p_bar_voltage} and \eqref{eq:p_bar_current}, one needs the open-circuit voltage $v_{oc}$, the voltage limits $\overline{v}, \underline{v}$, the rated current $\overline{i}$ and the series resistance $R$.
While the rated current, voltage limits and series resistance are constant and known, the open-circuit voltage depends on the battery state of charge, which depends, in turn, on the battery charging/discharging power. Therefore, as shown next, Eq.~\eqref{eq:dpc_max} features an interdependency between the current and past control actions that must be explicitly modeled.

\subsection{Power constraints as a function of the battery SOC}
\begin{figure}[!h]
\centering
\includegraphics[width=1\columnwidth]{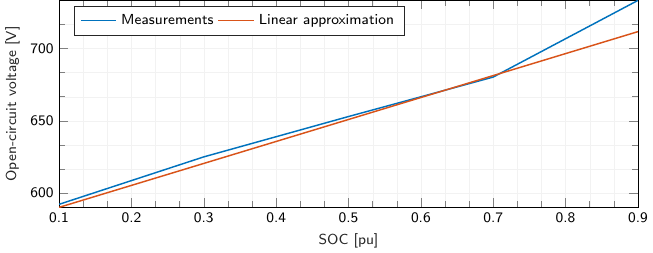}
\caption{Open-circuit voltage as a function of the battery state of charge for the lithium-ion BESS considered in this paper's case study.}\label{fig:soc-to-voltage}
\end{figure}

The open-circuit voltage of a battery depends on its SOC, as shown in Fig.~\ref{fig:soc-to-voltage} by the measurements from the lithium-ion BESS considered in this paper. We approximate the SOC-to-voltage relation with a linear model that, as Fig.~\ref{fig:soc-to-voltage} shows, holds well for SOC values in the central part of the SOC range.  Because BESS normally operates with some margin from the extreme SOC limits, this approximation is considered to hold reasonably well in practical applications.

\begin{figure}[!h]
\centering
\footnotesize
\includegraphics[width=1\columnwidth]{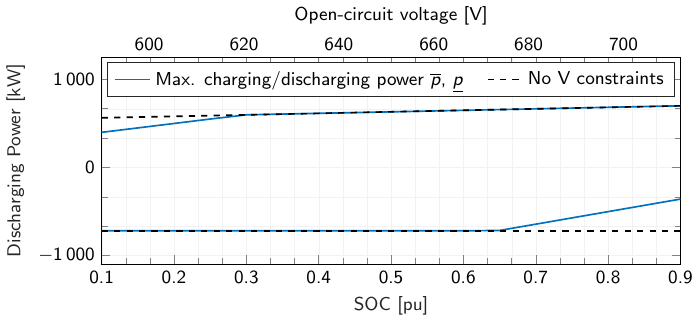}
\caption{The feasible power area of the BESS as a function of the battery SOC considering both voltage and current constraints (blue lines) and without voltage constraints (dashed black lines)}\label{fig:dpc}
\end{figure}

The linear SOC-to-voltage approximation of Fig.~\ref{fig:soc-to-voltage} is now used to calculate the power bounds \eqref{eq:p_bar_voltage}-\eqref{eq:dpc_max} across the whole SOC range. This result is shown in Fig.~\ref{fig:dpc}: the envelope between the two solid blue curves is the feasible operating range of the battery power; by construction, it denotes the surface where current and voltage constraints are respected. Fig.~\ref{fig:dpc} shows two features: first, the discharging power monotonically increases with the battery SOC (i.e., the battery can provide more power when fully charged than when discharged) and vice-versa for the charging power (i.e., at high SOC values, the battery can recharge at a lower rate than when discharged). Second, the feasible operating area is a convex set and can thus be represented with linear inequalities. We prove in Appendix A that the linear relationship between the battery open-circuit voltage and SOC is a sufficient condition to ensure the convexity of the envelope in Fig.~\ref{fig:dpc}.

In addition, Fig.~\ref{fig:dpc} shows the feasible power region of the BESS without considering the voltage constraints (as, for example, done in \citep{8770143}) with dashed black lines: these lines diverge from the blue solid curves at low and high SOC levels, denoting that, for low and high SOC values, voltage limits are more binding on the BESS power than current constraints.

Now that a feasible operating range for the BESS power has been found, it can be included in the scheduling problem, as shown in the next subsection.

\subsection{Inclusion of power constraints in the scheduling problem}\label{sec:main_result}
The power capabilities curves of Fig.~\ref{fig:dpc}, which denote the feasible operating region for the battery power, are used to replace the static power constraints \eqref{eq:powerconstraint0} in the scheduling problem. 
We model the upper and lower bounds of the convex envelope of Fig.~\ref{fig:dpc} with the following equations, linear in the variable $SOC_t$:
\begin{subequations}
\begin{align}
\overline B_k = \overline a_k + \overline b_k \cdot SOC_t, && k=1, \dots, K \\
\underline B_j = \underline a_j + \underline b_j \cdot SOC_t && j=1, \dots, J 
\end{align}\label{eq:linear_envelope}\end{subequations}
where $\left(\overline{a}_k,\underline{a}_j\right)$ and $\left(\overline{b}_k,\underline{b}_j\right)$ are coefficients of the linear functions estimated from the curves, and $K$ and $J$ are the numbers of linear equations used to approximate the envelope.

By replacing \eqref{eq:soc} into \eqref{eq:linear_envelope}, one can write the BESS power constraints in the following form:
\begin{align}
B_t \leq \overline a_k + \overline b_k \cdot \left[ SOC_{0} - h\left(\seq{B}{0}{t-1}\right) \right], && \text{for all $k$} \\
B_t \geq \underline a_j + \underline b_j \cdot \left[ SOC_{0} - h\left(\seq{B}{0}{t-1}\right) \right], && \text{for all $j$}
\end{align} 
which highlights a dependency between the current battery power $B_t$ and its history $B_0, B_1, \dots, B_{t-1}$. Finally, the scheduling problem presented in \eqref{eq:scheduler0} can be reformulated as follows:
\begin{subequations}\label{eq:scheduler1}
\begin{align}
 \underset{\seq{F}{0}{T}}{\text{arg~min}} \left\{ \sum_{t=0}^{T} f(F_t) \right\}
\end{align}
subject to
\begin{align}
    & SOC_{t+1} = SOC_0 - h\left(\seq{B}{0}{t}\right) \\
    & \underline{SOC} \le SOC_{t+1} \le \overline{SOC}
\end{align}
for $t=0,1,\dots,T$, and    
\begin{align}
    & B_t \leq \overline a_k + \overline b_k \cdot \left[ SOC_{0} - h\left(\seq{B}{0}{t-1}\right) \right] \\ 
    & B_t \ge \underline a_j + \underline b_j \cdot \left[ SOC_{0} - h\left(\seq{B}{0}{t-1}\right) \right]
\end{align}\end{subequations}
for $t=0,1,\dots,T$, $k=1,2,\dots,K$, and $j=1,2,\dots,K$. It is important to remark that if the function $h_{[t]}(\cdot)$ can be reformulated as a linear function (by way of integer variable or convex relaxations, e.g., \citep{boyd}), the set of constraints become linear in the battery power $B_t$, thus leaving unaltered the original convexity property of the problem.  We refer to this set of constraints as DPC.

Finally, it is interesting to remark that if the BESS internal resistance changes due to, for example, electrochemical degradation, the feasible power area of the BESS and the DPC can be recomputed by updating the value of $R$ in the computation of the power constraints.

\section{Application Example: Hybrid Hydropower Plant}\label{sec:appl}
This section describes a practical implementation of the dynamic power constraints proposed in this paper. In particular, it formulates a scheduling problem for a power-intensive application that serves as a benchmarking ground to compare the performance of static and dynamic power constraints. Results will be then presented and evaluated in Section~\ref{sec:results}. 

It is important to remark that the contributions of this paper on the BESS power constraints for scheduling problems are independent from this specific application example and can be generalized to other BESS use cases, such as provision of ancillary services, electricity market trading, and energy management.

As a power-intensive application, we select a hybrid HPP, which, as will be evident in the formulation, demands substantial levels of power and relatively small values of energy from BESSs.

\subsection{Hybrid hydropower plant}
For the unfamiliar reader, a hybrid HPP is a conventional HPP coupled with a BESS installed at the plant's premises. The role of the BESS is to take over those power regulation duties from the HPP power setpoints, such as fast transients, that engender excess mechanical fatigue on the HPP components so as to avoid premature mechanical failures. The HPP and BESS are operated by a real-time controller, which decides how to split the power among them. Multiple real-time controllers have been proposed in the literature, as a function of the adopted methodology and type of HPP (e.g., \citep{GERINI2021100538,en16135074,10106212,10565982}). In this paper, we use the model predictive control proposed in \citep{Cassano2021,CASSANO2022108545} that aims to limit the pressure transients on the penstock caused by the water-hammer effect in high- and medium-head hydropower plant.


\begin{figure*}[ht!]
    \centering
    {\footnotesize
\sffamily
    \includegraphics[width=1.8\columnwidth]{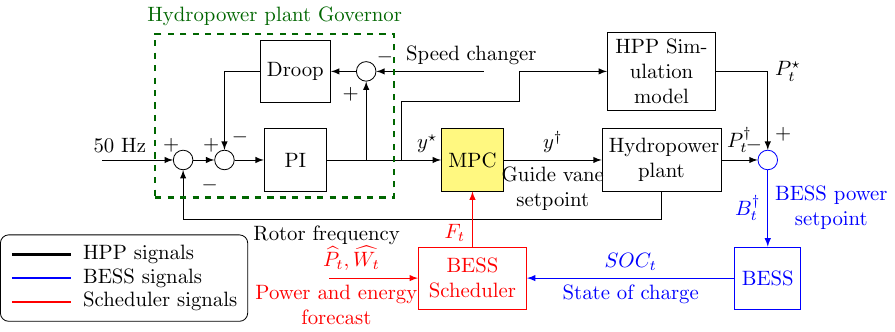}
    }
    \caption{The control scheme of a hybrid hydropower plant.}
    \label{fig:hhpp}
\end{figure*}

The components of the real-time control adopted for the selected use case are shown in Fig.~\ref{fig:hhpp}. The HPP is equipped with a conventional plant governor for primary and secondary frequency regulation. The governor provides a reference position $y_t^\star$ for the guide vane, that is the organ that adjusts the water flow and regulates the power output.  $y_t^\star$ is processed by the MPC \citep{Cassano2021} that reduce mechanical fatigue and produces a new guide vane setpoint $y_t^\dagger$ that is sent for actuation. The BESS power setpoint is computed so that the hybrid power plant, globally, maintains the same power output of the HPP as if it were controlled without the filtering action of the MPC; it is calculated as the difference between the HPP power output without the MPC action, $P_t^\star$, and the HPP power output without MPC, $P_t^\dagger$.

The real-time control operates at a fast pace (subsecond resolution) and computes the BESS and hydropower plant's setpoint. As a result of the actuation of the BESS setpoints, the battery SOC will unavoidably drift from its initial value. In order to compensate for this drift and maintain an adequate SOC level, a scheduler is necessary.

\subsection{BESS scheduler}
The BESS scheduler in the hybrid HPP ensures that the real-time control has access to sufficient power and energy in the BESS to provide the prescribed fatigue-reduction services. The BESS scheduler constantly monitors the SOC of the BESS and re-computes periodically, each 90 seconds, a secondary setpoint for the BESS so as to steer it to a feasible state.

Three models are instrumental in formulating the schedulers: power and energy forecasts to compute the prediction of the battery SOC, the SOC constraints, and the battery power constraints formulated with the method described in Section~\ref{sec:DPCs}.

\subsubsection{BESS power and SOC evolution}
As introduced for \eqref{eq:spc}, the BESS power can be modeled as:
\begin{align}
    \widehat{B}_t = \widehat{P_t} + F_t, && t=0,\dots,T-1 \label{eq:b_hat}
\end{align}
where $\widehat{P_t}$ is a power forecast of the service to provide, and $F_t$ is offset profile. As the power $\widehat{P_t}$ of the service to provide might feature large volatility, as for primary frequency control, it is difficult to be forecasted as a point prediction (i.e., the expected value of the realization). For this reason, it is modeled in terms of Prediction Intervals (PI), which provide a range where the realization is expected to lay with a given confidence level.
We denote the lower and upper bound of the PI of $P_t$ with $\widehat{P}^{\downarrow}_t$ and $\widehat{P}^{\uparrow}_t$, respectively, with $\widehat{P}^{\downarrow}_t \le \widehat{P}^{\uparrow}$.
Replacing these forecasts into \eqref{eq:b_hat} yield the power bounds for the BESS:
\begin{subequations}\label{eq:B_hat}
\begin{align}
\widehat{B}^{\uparrow}_t = \widehat{P}^{\uparrow}_t + F_t \\
\widehat{B}^{\downarrow}_t = \widehat{P}^{\downarrow}_t + F_t.
\end{align}
\end{subequations}

By replacing the predictions \eqref{eq:B_hat} into the state of charge model in \eqref{eq:soc}, one could obtain PIs of the battery SOC under these operating conditions. However, static PIs in the SOC model might lead to unreasonably large energy needs as this entails integrating a constant value over time; in order to avoid large energy needs in the SOC model, we introduce a second set of PIs, called $\widehat{W}^{\uparrow}_t, \widehat{W}^{\downarrow}_t$ (assumed with smaller absolute value than $\widehat{P}^{\downarrow}_t, \widehat{P}^{\uparrow}_t$), which refers to the energy forecasts for the service to deliver, and use these in the SOC model to obtain less conservative estimates of SOC needs. To exemplify the situation, let us consider the provision of primary frequency regulation with a droop controller. The power to deliver is proportional to the deviation of the grid frequency from the nominal value (50 Hz), the droop coefficient, and the plant's rated power. This power contribution for primary frequency control might be large; however, it is not sustained for long because, in minutes, secondary frequency control intervenes to re-dispatch generators, releasing the power previously activated for primary frequency control. 
Thus, being primary frequency control characterized by potentially large power values but modest energy, it is a power-intensive application. Power forecasts $\widehat{P}^{\downarrow}_t, \widehat{P}^{\uparrow}_t$ and energy forecasts $\widehat{W}^{\uparrow}_t, \widehat{W}^{\downarrow}_t$ are designed with the spirit of capturing this property. An example of the computation of these forecasts is given in Section~\ref{sec:case}.

Energy PIs are applied to the model \eqref{eq:soc} to estimate the PIs of the SOC. This reads as:
\begin{subequations}
\begin{align}
\widehat{SOC}^{\downarrow}_{t+1} = SOC_{0} - h\left(\seq{\widehat{W}^{\uparrow} + F}{0}{t}\right) \\
\widehat{SOC}^{\uparrow}_{t+1} = SOC_{0} - h\left(\seq{\widehat{W}^{\downarrow} + F}{0}{t}\right),
\end{align}\label{eq:SOC_hat}
\end{subequations}
for all $t$. These estimations will be used in the next paragraph to formulate constraints on the battery SOC. 

\subsubsection{SOC constraints}
The battery SOC in \eqref{eq:SOC_hat} should remain within (customizable) limits $\underline{{SOC}}, \overline{{SOC}}$. Because $\widehat{{SOC}}^{\uparrow}_{t}$ is larger than $\widehat{{SOC}}^{\downarrow}_{t}$ by construction, we can apply the upper bound to $\widehat{{SOC}}^{\uparrow}_{t}$ and the lower bound to $\widehat{{SOC}}^{\downarrow}_{t}$:
\begin{align}
\widehat{SOC}^{\uparrow}_{t} \le \overline{{SOC}} \\
\widehat{SOC}^{\downarrow}_{t} \ge \underline{{SOC}},
\end{align}
for all $t$.

\subsubsection{BESS power constraints}
We are concerned with ensuring that the BESS expected power, modeled in \eqref{eq:B_hat} with PIs, does not exceed the BESS power capabilities, which, as explained in \ref{sec:main_result}, depend on the battery SOC. Recalling from the previous section, the battery power limits are given by:
\begin{subequations}\label{eq:recall}
\begin{align}
& \overline B \le \overline a_k + \overline b_k \cdot SOC_t, && k=1, \dots, K \\
& \underline B \ge \underline a_j + \underline b_j \cdot SOC_t && j=1, \dots, J 
\end{align}
\end{subequations}
where $\overline a_k$, $\overline b_k$, $\underline a_j$, and $\underline b_j$ are given parameters, and $SOC_t$ is the state of charge at the time interval of interest (which in turn depends on turn on the battery utilization history).

Because $\widehat{B}^{\uparrow}_t > \widehat{B}^{\downarrow}_t$ by construction, one could proceed with applying the larger power limit $\overline B$ to the battery power's upper bound $\widehat{B}^{\uparrow}_t$, and the lower limit $\underline B$ to the battery power's lower bound $\widehat{B}^{\downarrow}_t$, (i.e., two sets of constraints). However, the quantity $SOC_t$, required in \eqref{eq:recall} to compute the bounds, is uncertain as it depends on the battery utilization. This uncertain SOC trajectory was characterized in \eqref{eq:SOC_hat} in terms of PIs with the identification of the upper and lower bound trajectories of the state of charge, $\widehat{SOC}^{\uparrow}_{t}$ and $\widehat{SOC}^{\downarrow}_{t}$, respectively. Because the trajectory of the SOC realization $SOC_t$ is likely to lie between $\widehat{SOC}^{\uparrow}_{t}$ and $\widehat{SOC}^{\downarrow}_{t}$, the battery power limits should be written for both these SOC trajectories in order to ensure that they are respected under any possible realization of the battery power within the identified PIs. This requirement leads to formulating two sets of constraints for the battery's power upper bound (one evolving with $\widehat{SOC}^{\uparrow}_{t}$ and the other with $\widehat{SOC}^{\downarrow}_{t}$), and two sets for the battery's power upper bound (i.e., four sets of constraints, in total). Finally, this formulation reads as:
\begin{subequations}\label{eq:b:constraints}
\begin{align}
& \widehat B^\uparrow_t \leq \overline a_k + \overline b_k \cdot \left[ SOC_{0} - h\left(\seq{\widehat{W}^{\uparrow} + F}{0}{t}\right) \right] \\
& \widehat B^\uparrow_t \leq \overline a_k + \overline b_k \cdot \left[ SOC_{0} - h\left(\seq{\widehat{W}^{\downarrow} + F}{0}{t}\right) \right] \\
& \widehat B^\downarrow_t \geq \underline a_j + \underline b_j \cdot \left[ SOC_{0} - h\left(\seq{\widehat{W}^{\downarrow} + F}{0}{t}\right) \right] \\
& \widehat B^\downarrow_t \geq \underline a_j + \underline b_j \cdot \left[ SOC_{0} - h\left(\seq{\widehat{W}^{\uparrow} + F}{0}{t}\right) \right].
\end{align}
\end{subequations}

\subsection{Formulation of the scheduling problem}
Consider to be at the time interval $t=0$; according to the estimation currently delivered by the battery management system, the BESS state of charge is $SOC_0$. The sequences $[P]_0^{(T)}$ and $[W]_0^{(T)}$ store, respectively, the forecasts for peak power utilization and energy utilization for the service that the BESS should deliver (e.g., fatigue reduction service in the hybrid HPP or primary frequency control) over the future time horizon $t=0,1,\dots,T$.

The BESS scheduling problem computes the sequence $[F]^{(T)}_0$ so as to ensure that the BESS constraints (state of charge $SOC_t$ and power $B_t$ for $t=0,\dots,T$) will be respected under the BESS utilization forecasts. The formulation reads as the following constrained optimization problem:
\begin{subequations}\label{eq:scheduler_app}
\begin{align}
 \underset{\seq{F}{0}{T}}{\text{arg~min}} \left\{ \sum_{t=0}^{T} F^2_t \right\} \label{eq:opt:f}
\end{align}
subject to the SOC models and constraints
\begin{align}
SOC_{0} - h\left(\seq{\widehat{W}^{\uparrow} + F}{0}{t}\right) \geq \underline{SOC} \\
SOC_{0} - h\left(\seq{\widehat{W}^{\downarrow} + F}{0}{t}\right) \leq \overline{SOC}
\end{align}
for $t=0,\dots,T$; and the battery power constraints
\begin{align}
& \widehat P^\uparrow_t + F_t \leq \overline a_k + \overline b_k \cdot \left[ SOC_{0} - h\left(\seq{\widehat{W}^{\uparrow} + F}{0}{t}\right) \right] \label{eq:opt:f:p0}\\
& \widehat P^\uparrow_t + F_t \leq \overline a_k + \overline b_k \cdot \left[ SOC_{0} - h\left(\seq{\widehat{W}^{\downarrow} + F}{0}{t}\right) \right] \\
& \widehat P^\downarrow_t + F_t \geq \underline a_j + \underline b_j \cdot \left[ SOC_{0} - h\left(\seq{\widehat{W}^{\downarrow} + F}{0}{t}\right) \right] \\
& \widehat P^\downarrow_t + F_t \geq \underline a_j + \underline b_j \cdot \left[ SOC_{0} - h\left(\seq{\widehat{W}^{\uparrow} + F}{0}{t}\right) \right], \label{eq:opt:f:pf}
\end{align}\end{subequations}
for $t=0,\dots,T$, $k=1,\dots,K$, and $j=1,\dots,J$.

The cost function in \eqref{eq:opt:f} is designed to minimize the norm-2 of $[F]_0^{(t)}$, which consists of the difference between the battery power output and its forecast (tracking error). The sequence $[F]_0^{(t)}$ will be of all zeros unless the BESS utilization forecasts will drive the battery out of its constraints.

Compared to the constraint set in \eqref{eq:b:constraints}, the left hand-sides of inequalities \eqref{eq:opt:f:p0}-\eqref{eq:opt:f:pf} have been rendered by using \eqref{eq:B_hat} to highlight the dependency of the constraints on the problem variable $F_t$.

The optimization problem \eqref{eq:scheduler_app} is applied in a receding horizon manner, i.e., the optimization is repeated periodically each 90~seconds with updated information; at each period, only the first decision action of the sequence $[F]$ is actuated. The (arbitrary) 90-second period is chosen empirically according to the input time series in order to restore a suitable SOC of the BESS.

As a final note, the formulation \eqref{eq:scheduler_app} features hard constraints. In practice, it is recommended to render these constraints as soft ones by adding slack variables to the problem. This will make it possible for the optimization problem to converge even when the hard constraints are violated, enabling the provision of the prescribed power and services in a best-effort mode with feasible control actions.

For the sake of completeness, it is interesting to show the implementation of the same scheduler as in \eqref{eq:scheduler_app} but with conventional static power constraints (SPC) instead of DPCs:
\begin{subequations}\label{eq:scheduler_app_conv}
\begin{align}
 \underset{\seq{F}{0}{T}}{\text{arg~min}} \left\{ \sum_{t=0}^{T} F^2_t \right\} \label{eq:opt:f}
\end{align}
subject to the SOC models and constraints
\begin{align}
SOC_{0} - h\left(\seq{\widehat{W}^{\uparrow} + F}{0}{t}\right) \geq \underline{SOC} \\
SOC_{0} - h\left(\seq{\widehat{W}^{\downarrow} + F}{0}{t}\right) \leq \overline{SOC}
\end{align}
for $t=0,\dots,T$; and the battery power constraints
\begin{align}
& \widehat P^\uparrow_t + F_t \leq B^r \\
& \widehat P^\downarrow_t + F_t \geq -B^r
\end{align}\end{subequations}
for $t=0,\dots,T$ and where $B^r$ is the rated power of the BESS.
As visible by comparing the two formulations, upgrading from static to dynamic power constraints requires replacing a set of linear constraints with another set of linear constraints.

\subsection{Case study} \label{sec:case}
The simulation case study is a 230~MW medium-head HPP hybridized with a 720~kVA/500~kWh Li-ion BESS providing enhanced grid balancing services to the grid in view of an increased proportion of production from renewable energies. The HPP has a net head of 315 meters. It is equipped with a Francis turbine and is fed by an open-air 1100-meter-long penstock. HPP and BESS characteristics are summarized in Table~\ref{tab:hpp}. The energy capacity and power rating of the BESS were selected this way because it was empirically verified that these levels of power and energy were sufficient to mitigate the power transients causing excess mechanical fatigue to the hydropower plant. Formal sizing of the BESS is beyond the objectives of the scheduling problem and will be investigated in future works. 

\begin{table}[ht!]
    \centering
    \caption{Parameters of the simulation case study}
    \includegraphics[width=0.9\columnwidth]{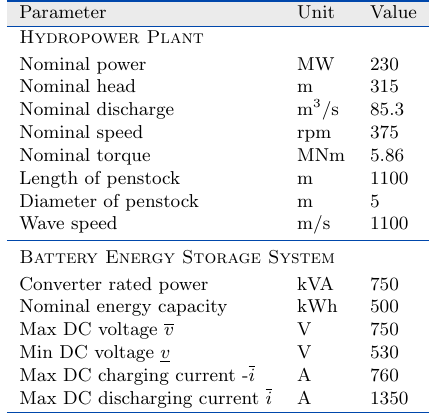}
\label{tab:hpp}
\end{table}

The HPP is simulated with an equivalent circuit model, as in \citep{Nicolet:98534}, that captures hydraulic transients, pressure losses in the pipes with one-dimensional equivalent circuit models, and turbine response through the characteristic curves. The synchronous generator torque is modeled with a second-order model, with the electrical torque as a function of the generator's power angle and rotor dynamics simulated with the swing equation.

The HPP governor consists of a standard Proportional-Integral (PI) governor with gains determined using the Ziegler-Nichols method and validated against the ENTSOE qualification tests for primary and secondary frequency control \citep{landry_methodology_nodate, Test}. The governor implements limits on the rate of change and the magnitude of the guide vane. In addition, it implements a speed droop and speed changer to supply primary and secondary frequency regulation to the grid. 

The speed droop coefficient is set to 2\%; compared to typical values (e.g., 5\%), a smaller droop coefficient is reproduce future operational settings where residual dispatchable generation assets, such as HPPs, provide higher flexibility to compensate for the missing regulation capacity after replacing conventional generation with renewables.

\begin{figure}[ht!]
    \centering
    \includegraphics[width=1\columnwidth]{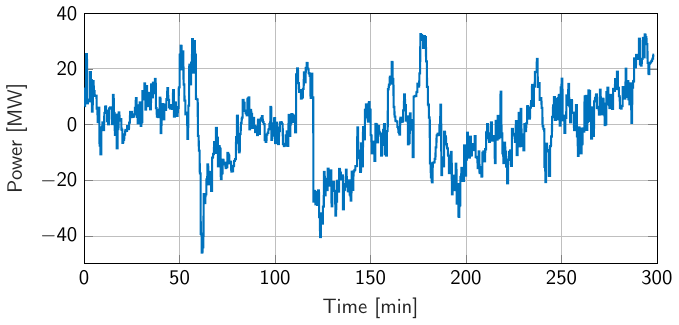}
    \caption{Power setpoints for secondary frequency control.}
    \label{fig:sfr}
\end{figure}

The power grid is modeled as an infinite bus, where it is the grid to impose the frequency under the assumption that its size is significantly larger than the plant.  Grid frequency in the simulations is reproduced using real system frequency measurements of the European interconnected system from \citep{EPFL-CONF-203775}; a portion of the grid frequency measurements with larger fluctuations than usual was selected by visual inspections so as to reproduce more intense grid balancing duties, thus indirectly reflecting higher variability of stochastic renewable sources and demand, and possibly reduced system inertia (see Fig.\ref{fig:sfr}).



\subsection{Forecasting BESS's power and energy needs for fatigue reduction}\label{sec:quantifying}
\begin{figure}[ht!]
    \centering
    \includegraphics[width=1\columnwidth]{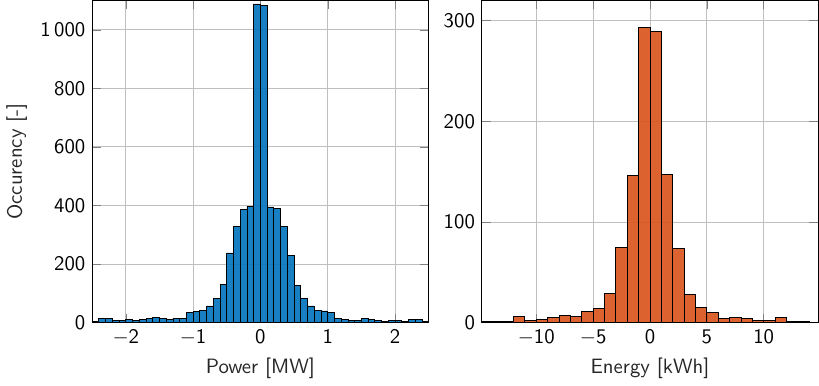}
    \caption{Empirical distributions of the battery power time series: original series sampled at a 1-second resolution (left), and resampled at a 10-second resolution by sample average (right).}
    \label{fig:distributions}
\end{figure}

The scheduling problem requires PIs of the power and energy of the BESS, which are estimated on the basis of historical data of the battery utilization. For the power PIs, an empirical probability distribution function (PDF) is built by computing the histogram historical BESS power time series at a 1-second resolution, shown in the left-panel plot of Fig.~\ref{fig:distributions}; PIs are then estimated as the 5\% and 95\% percentiles of such a PDF.  

As far as energy PIs are concerned, the BESS power time series at a 1-second resolution is resampled to 90 seconds by sample average and multiplied for the new time resolution to obtain energy values; the histogram of the resampled time series, shown in the right-panel plot of Fig.~\ref{fig:distributions}, is used to estimate its empirical PDF. Energy PIs are finally estimated by calculating the 5\% and 95\% percentiles of this PDF.

By comparing the power and energy histograms in Fig.~\ref{fig:distributions}, one can note the power-intensive nature of the considered service, with power values in the order of magnitude of MW and energy values in the order of kWh.

\section{Results and Discussion} \label{sec:results}
This section scrutinizes the performance of the scheduler equipped with DPC and compares it against two others: conventional SPC, and DPC without voltage constraints.

Performance is evaluated by quantifying the current violations that occur during BESS real-time operations when deploying the three schedulers. As this section will show, the scheduler with DPC achieves fewer and smaller current constraint violations than SPC and DPC without voltage constraints.

The simulation case study is as discussed in Section~\ref{sec:case}; the lower-level control is actuated each second; the scheduler is updated in a receding horizon fashion every 90 seconds. Control and scheduling setpoints are actuated in the simulation model described in Section~\ref{sec:case}.

\begin{figure}[!ht]
	\centering
		\includegraphics[width=1\columnwidth]{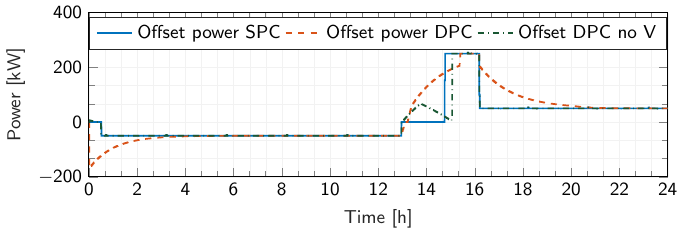}
  \\
		\includegraphics[width=1\columnwidth]{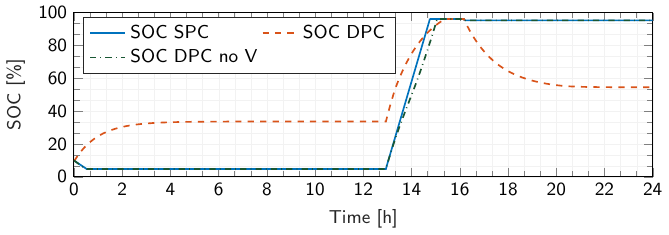}
	\caption{Offset power $[F]$ computed by the scheduler with SPC, DPC and DPC without voltage constraints (top panel) and battery SOC evolution (lower panel).}
	\label{fig:sched_comp}
\end{figure}

Fig.~\ref{fig:sched_comp} shows the offset profiles with SPC, DPC and DPC without voltage constraints (upper-panel plot) and the SOC evolution (lower-panel plot), and it will now be explained. In this simulation, the BESS is operated near either the low SOC values or high SOC values to reproduce challenging conditions for the scheduler. This operating condition could reflect, for example, the provision of multiple services in addition to the fatigue reduction service. The power and energy prediction intervals are $\hat{P}^{\uparrow}_t = 600$ kW, $\hat{P}^{\downarrow}_t = -600$ kW, $\hat{W}^{\uparrow}_t = 4.1$ kWh and $\hat{W}^{\downarrow}_t = -5.2$ kWh (calculated as described for Fig.~\ref{fig:distributions}). 

Fig.~\ref{fig:sched_comp} shows that the DPC schedule features an initial large charging requirement that brings the battery SOC to 40\%. This is because, as per forecasts, the battery cannot provide the forecasted discharging power between $\pm$600~kW at such a low SOC. At $t=12$~h, the battery is requested to provide a charging power of 250~kW for 3 hours and 50 kW for 9 hours by the additional service. The scheduler with DPC generates a positive offset profile to discharge the battery and keep the SOC level around 60 \%, where the battery can deliver the predicted power.  In contrast, the scheduler with SPC produces an offset profile that keeps the SOC within the limits of 5\%-95\%, as at these levels, the static power constraints allow the predicted power to be delivered.

The scheduler with DPC without voltage constraints behaves similarly to the one with SPC, maintaining the battery at the lower SOC limit of 5\%. At $t=13$~h, it charges the battery slightly faster than the SPC-based scheduler but still respects the 95\% SOC limit.




\subsubsection{Assessment of violations of current constraints}

\begin{figure*}[ht!]
    \centering
    \includegraphics[width=1\textwidth]{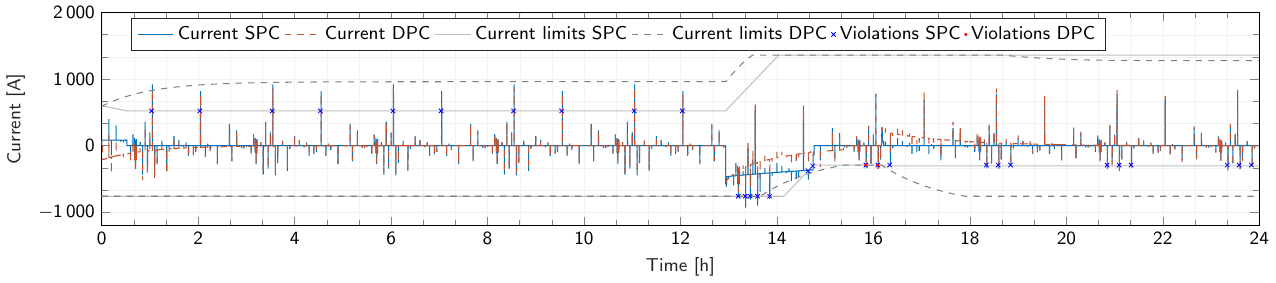}
    \caption{Battery currents, current limits and violations under the scheduler with static and dynamic power constraints.}
    \label{fig:curr_viol}
\end{figure*}

Fig.~\ref{fig:curr_viol} displays the BESS output current resulting from the DPS and SPC schedulers' actions and the current limits that are a function of the SOC of the battery. The current is retrieved from the BESS power and SOC, leveraging the battery model presented in Fig.~\ref{fig:batt_circuit}. In the first 12 hours, the scheduler with DPC charges the battery, increasing the charging current to 960~A. In contrast, the scheduler with SPC maintains a constant SOC of 5\%, corresponding to a charging current limit of 520~A. 
Due to this control action, the number of current limit violations, indicated by the blue and red markers in Fig.~\ref{fig:curr_viol}, is 10 for the scheduler with SPC and 0 for the scheduler with DPC. During the remaining simulation period, the scheduler with DPCs increases the discharging current capability to a value of -760~A while the scheduler with SPCs maintains the limit of -304~A. This leads the scheduler with SPCs to violate the discharging current limits 19 more times against two times from the scheduler with DPCs.  It is worth noting that the scheduler with DPC also exceeds current limits, which is related to the accuracy of the power and energy forecasts.

\begin{figure*}[ht!]
    \centering
    \includegraphics[width=1\textwidth]{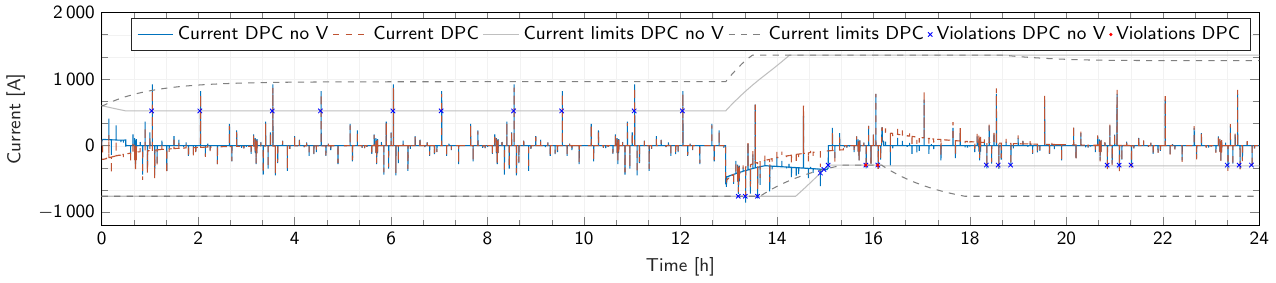}
    \caption{Battery currents, current limits and violations under the scheduler with dynamic power constraints and the one without voltage constraints.}
    \label{fig:curr_viol_noV}
\end{figure*}

Fig.~\ref{fig:curr_viol_noV} shows the BESS output current resulting from the scheduler with DPC and the DPC without voltage constraints. The total number of current violations for the scheduler with DPC is 2, while for the scheduler without voltage constraints is 27.

\subsubsection{Performance analysis for different initial SOC levels}
The analysis has been repeated for different initial SOCs. For initial SOCs between 10 and 50\%, it is considered the same additional service and prediction intervals as in the previous analysis, while for SOCs between 60\% to 90\%, the battery must provide a constant discharging power of 50 kW for 12 hours, 250kW of charging power for 3 hours and 50 kW of charging power for the remaining 9 hours. In the latter case, the power predictions remain the same, and the energy predictions are: $\hat{W}^{\uparrow}_t = 5.2$ kWh and $\hat{W}^{\downarrow}_t = -4.1$ kWh.

The three schedulers are compared based on these metrics: the number of current limit violations, the average difference of the current peaks between the upper and lower current limits, and their variance. The results are summarized in Table~\ref{tab:sched_stat}; the following conclusions can be drawn:
\begin{enumerate}
    \item the scheduler with DPC reduces the number of current violations by 93\% in the range 10-50\% and by 85\% in the range 60-90\% of the initial SOC;
    \item the scheduler with DPC achieves smaller current constraint variations, with a maximum reduction of 60\% of the average difference of the current peaks from the limits and a variance lower by two orders of magnitudes;
    \item the scheduler with DPC without voltage constraints performs similarly to the one with SPC.
\end{enumerate}
In conclusion, the results demonstrate that:
\begin{enumerate}
    \item implementing SOC-dependent power constraints in the scheduler's problem formulation reduces the probability of producing unfeasible schedules in power-intensive applications;
    \item including voltage constraints in the DPS significantly impacts the scheduler's performance.
\end{enumerate}


\begin{table}[!ht]
\renewcommand{\arraystretch}{1.2}
\centering
\caption{Schedulers' perfomance comparison on current limits violation.}\label{tab:sched_stat}
\begin{adjustbox}{width=1\columnwidth,center}
\begin{tabular}{c c c c c c C{1.2cm}}
\hline
\rowcolor{tableheadcolor}
Type of &  N$^o$ of  &  Average diff. &  Variance &  Average diff. &  Variance &  Initial\\ \rowcolor{tableheadcolor}
 scheduler &  violations &  (upper) [A] &  (upper) & (lower) [A] &  (lower) &  SOC [\%]\\
\hline \addlinespace[0.3ex]
SPC & 29 & 546.63 & 2.52$\cdot$10$^3$ & -93.51 & 7.82$\cdot$10$^3$ \\ 
DPC (no V) & 27 & 546.63 & 2.52$\cdot$10$^3$ & -74.43 & 5.62$\cdot$10$^3$ & \multirow{1}{1.2cm} {\centering 10\%}\\ 
DPC & 2 & - & - & -53.35 & 13.72  &\\
\hline  \addlinespace[0.3ex]
SPC & 28 & 546.63 & 2.52$\cdot$10$^3$ & -93.50 & 7.81$\cdot$10$^3$ \\ 
DPC (no V) & 26 & 546.63 & 2.52$\cdot$10$^3$ & -74.43 & 5.62$\cdot$10$^3$ & \multirow{1}{1.2cm} {\centering 20\%}\\ 
DPC & 2 & - & - & -53.35 & 13.72  &\\
\hline  \addlinespace[0.3ex]
SPC & 28 & 546.62 & 2.52$\cdot$10$^3$ & -93.48 & 7.81$\cdot$10$^3$ \\ 
DPC (no V) & 26 & 546.63 & 2.52$\cdot$10$^3$ & -74.43 & 5.62$\cdot$10$^3$ & \multirow{1}{1.2cm} {\centering 30\%}\\ 
DPC & 2 & - & - & -53.35 & 13.72  &\\
\hline \addlinespace[0.3ex]
SPC & 28 & 546.62 & 2.52$\cdot$10$^3$ & -93.48 & 7.81$\cdot$10$^3$ \\ 
DPC (no V) & 26 & 546.63 & 2.52$\cdot$10$^3$ & -74.43 & 5.62$\cdot$10$^3$ & \multirow{1}{1.2cm} {\centering 40\%}\\ 
DPC & 2 & - & - & -53.35 & 13.72  &\\
\hline  \addlinespace[0.3ex]
SPC & 28 & 546.61 & 2.52$\cdot$10$^3$ & -93.48 & 7.81$\cdot$10$^3$ \\ 
DPC (no V) & 26 & 546.63 & 2.52$\cdot$10$^3$ & -74.43 & 5.62$\cdot$10$^3$ & \multirow{1}{1.2cm} {\centering 50\%}\\ 
DPC & 2 & - & - & -53.34 & 3.86$\cdot$10$^3$  &\\
\hline  \addlinespace[0.3ex]
SPC & 21 & 335.06 & 2.06$\cdot$10$^3$ & -83.46 & 488.26 \\ 
DPC (no V) & 19 & 267.17 & 1.87$\cdot$10$^3$ & -83.46 & 488.26 & \multirow{1}{1.2cm} {\centering 60\%}\\ 
DPC & 3 & 134.65 & 43.13 & - & - &\\
\hline  \addlinespace[0.3ex]
SPC & 21 & 335.06 & 2.06$\cdot$10$^3$ & -83.48 & 488.26 \\ 
DPC (no V) & 19 & 267.17 & 1.87$\cdot$10$^3$ & -83.46 & 488.26 & \multirow{1}{1.2cm} {\centering 70\%}\\
DPC & 3 & 134.65 & 43.13 & - & - &\\
\hline \addlinespace[0.3ex]
SPC & 21 & 335.08 & 2.06$\cdot$10$^3$ & -83.48 & 488.27 \\
DPC (no V) & 19 & 267.17 & 1.87$\cdot$10$^3$ & -83.46 & 488.26 & \multirow{1}{1.2cm} {\centering 80\%}\\
DPC & 3 & 134.65 & 43.13 & - & - &\\
\hline  \addlinespace[0.3ex]
SPC & 22 & 335.08 & 2.07$\cdot$10$^3$ & -83.50 & 488.27 \\ 
DPC (no V) & 20 & 267.17 & 1.87$\cdot$10$^3$ & -83.46 & 488.26 & \multirow{1}{1.2cm} {\centering 90\%}\\
DPC & 3 & 134.65 & 43.13 & - & - &\\
\hline
\end{tabular}
\end{adjustbox}
\end{table}

\section{Conclusions} \label{sec:concl}

By leveraging an approach based on equivalent circuit models, this paper formulated a novel set of BESS power constraints for scheduling problems that are capable of more accurate estimates of voltage and current constraints compared to the existing literature. We have shown that these constraints are linear in the battery power, thus making it possible to retrofit existing scheduling problems without altering their tractability property (i.e., convexity). As an application example, we have considered a hybrid hydropower plant as a use case. In this context, we compared the traditional scheduler with static power constraints and with dynamic power constraints without considering voltage constraints against the proposed one based on (i) the production of a feasible schedule and (ii) the respect of the current limits, under different operating conditions. Results showed that the proposed scheduler performed better in both comparisons. Specifically, it produced a schedule that respected the SOC limitations and significantly reduced the number of current violations by 93\% in the SOC range of 10-50\% and by 85\% in the range of 60-90\%. Therefore, implementing SOC-dependent power constraints into the problem formulation of the scheduler is of utmost importance to optimize its efficacy in power-intensive applications. Future works refer to the validation of the proposed scheduler in an experimental setting.

\appendix

\section{Convexity of the power bounds}

\textbf{Statement.} \emph{A linear relation between the battery open-circuit voltage and state of charge is a sufficient condition to ensure a convex feasible region of the BESS power.}

We denote the epigraph and hypograph of the feasible set in \eqref{eq:dpc_max} and Fig.~\ref{fig:dpc} by $g_2$ and $g_1$, respectively. For the set to be convex, $g_2$ needs to be convex and $g_1$ concave. Because the composition of a composition of a convex (concave) function with a linear function preserves convexity (concavity, respectively), the claim is proved.




\bibliographystyle{elsarticle-num} 
\bibliography{biblio}

\end{document}